\documentclass[fleqn,usenatbib]{mnras}

\usepackage[T1]{fontenc}
\usepackage{ae,aecompl}
\usepackage[utf8]{inputenc}

\usepackage{graphicx}	
\usepackage{amsmath}	
\usepackage{amssymb}	
\usepackage{multirow}
\usepackage{txfonts}
\usepackage[utf8]{inputenc}
\usepackage{float}
\usepackage{subcaption}
\usepackage[T1]{fontenc}
\usepackage{microtype}
\usepackage{booktabs}
\usepackage{caption}
\usepackage{enumitem}

\usepackage{subfiles} 
\usepackage[dvipsnames]{xcolor}
\usepackage[decimalsymbol=., expproduct=times, separate-uncertainty=true, multi-part-units=single]{siunitx} 
\DeclareSIUnit\erg{erg}
\DeclareUnicodeCharacter{2010}{-}

\newcommand{\Go}{$\mathrm{G}_0\mathrm{\ }$} 
\newcommand{\HSM}{$\rho_{100} \mathrm{\ }$}
\newcommand{\FSM}{$\rho_{50} \mathrm{\ }$}
\newcommand{\rHSM}{$\rho \sim 100 \mathrm{\ M}_\odot \mathrm{\ pc}^{-3}\mathrm{\ }$}
\newcommand{\rFSM}{$\rho \sim 50 \mathrm{\ M}_\odot \mathrm{\ pc}^{-3}\mathrm{\ }$}

\title[External photoevaporation constrains planet formation]{External photoevaporation of circumstellar disks constrains the timescale for planet formation}

\author[Concha-Ramírez et al.]{Francisca Concha-Ramírez$^1$\thanks{E-mail: fconcha@strw.leidenuniv.nl},
Martijn J. C. Wilhelm$^1$,
Simon Portegies Zwart$^1$,\newauthor
Thomas J. Haworth$^{2, 3}$
\\
$^{1}$
Leiden Observatory, Leiden University, PO Box 9513, 2300 RA Leiden, The Netherlands\\
$^{2}$
Astrophysics Group, Imperial College London, Blackett Laboratory, Prince Consort Road, London SW7 2AZ, UK\\
$^{3}$
Astronomy Unit, School of Physics and Astronomy, Queen Mary University of London, London E1 4NS, UK
}

\date{Accepted XXX. Received YYY; in original form ZZZ}

\pubyear{2019}

\begin{document}
\label{firstpage}
\pagerange{\pageref{firstpage}--\pageref{lastpage}}
\maketitle

\begin{abstract}
Planet-forming circumstellar disks are a fundamental part of the star formation process. Since stars form in a hierarchical fashion in groups of up to hundreds or thousands, the UV radiation environment that these disks are exposed to can vary in strength by at least six orders of magnitude. This radiation can limit the masses and sizes of the disks. Diversity in star forming environments can have long lasting effects in disk evolution and in the resulting planetary populations. We perform simulations to explore the evolution of circumstellar disks in young star clusters. We include viscous evolution, as well as the impact of dynamical encounters and external photoevaporation. We find that photoevaporation is an important process in destroying circumstellar disks: in regions of stellar density \rHSM around $80\%$ of disks are destroyed before $\SI{2}{Myr}$ of cluster evolution. In regions of \rFSM around $50\%$ of disks are destroyed in the same timescale. Our findings are in agreement with observed disk fractions in young star forming regions and support previous estimations that planet formation must start in timescales $< 0.1 - 1 \mathrm{\ Myr}$.
\end{abstract}


\section{Introduction}
\label{sec:introduction}

Circumstellar disks develop as a result of the star formation process \citep{williams2011}. Since a non negligible fraction of stars are not born in isolation \citep{bressert2010,king2012}, and gas left over from the star formation process can linger for a few Myr \citep{portegieszwart2010}, during their first stages of evolution the disks remain embedded in an environment that is dense in gas and neighbouring stars. These conditions can be hostile for the disks in a myriad of ways: they can be subject to dynamical truncations \citep{vincke2015,portegieszwart2016,vincke2016} or be affected by processes related to stellar evolution, such as stellar winds \citep{pelupessy2012}, supernovae explosions \citep{close2017,portegieszwart2018}, and photoevaporation due to bright OB stars in the vicinity \citep[e.g.][]{guarcello2016,haworth2017}. The surrounding gas can also shrink the disks through ram pressure stripping \citep{wijnen2016, wijnen2017}. Since planet formation related processes seem to start very quickly in circumstellar disks ($< 0.1 - \SI{1}{Myr}$, \citet{najita2014,manara2018}), understanding the mechanisms that affect disk evolution is directly connected to understanding planetary system formation and survival. The Sun was probably born within a star cluster \citep{portegieszwart2009}, so discerning how the cluster environment affects the evolution of the disks can help us comprehend the origins of the Solar System. 

There are several observational indications that the environment of circumstellar disks shortly after their formation is unfavorable for their survival. Disks have been observed to be evaporating in several young star forming regions \cite[e.g.][]{fang2012,dejuanovelar2012,mann2014,kim2016,vanterwisga2019a}. Moreover, observations indicate that disk fractions decline in regions close to an O-type star \citep[e.g.][]{balog2007,guarcello2007,guarcello2009,guarcello2010,fang2012,mann2014,guarcello2016}. \citet{fatuzzo2008} estimate an FUV radiation field of up to $\mathrm{G}_0 \approx 1000$ in star clusters of $N > 1000$ stars\footnote{$\mathrm{G}_0$ is the FUV field measured in units of the Habing flux, $\SI{1.6e-3}{\erg\per\second\per\square\cm}$ \citep{habing1968}.}, while \citet{facchini2016} show that disks of radius $\sim\SI{150}{au}$ are subject to photoevaporation even in very low FUV fields ($\mathrm{G}_0 = 30$). In regions of high stellar density, nearby stars can also affect disk size and morphology by dynamical interactions. Observational evidence of the imprints of dynamical truncations has been reported in several nearby stellar clusters \citep{olczak2008,reche2009,dejuanovelar2012}. Tidal stripping that can be explained by disk-star interactions has been observed in the RW Aurigae system \citep{cabrit2006,rodriguez2018,dai2015} and in the T Tauri binary system AS 205 \citep{salyk2014}. There is also evidence that the Solar System might have been affected by a close encounter with another star during its early stages \citep{jilkova2015,pfalzner2018}.

Circumstellar disks are not only affected by external processes, but also by their internal viscous evolution. The combination of outwardly decreasing angular velocity together with outwardly increasing angular momentum causes shear stress forces inside the disks. As a consequence mass is accreted from the innermost regions of the disk onto its host star, whereas the outermost regions expand \citep{lynden-bell1974}. \citet{tazzari2017} propose that the measured offsets in sizes and masses of disks in the Lupus clouds versus disks in the Taurus-Auriga and Ophiuchus regions can be explained as observational evidence of viscous spreading. However \citet{rosotti2019} argue that current surveys do not yet have the sufficient sensitivity to properly detect this phenomenon.

Different approaches have been implemented to study the effects of these processes on the lifetime of circumstellar disks. External photoevaporation has been modeled with radiation hydrodynamics codes that solve flow equations through the disk boundaries, together with photodissociation region codes to obtain the temperature profiles of the disks \citep[e.g.][]{haworth2016,facchini2016}. This method has been coupled with $\alpha$-disk models to account for viscous evolution of the disk \citep[e.g.][]{adams2004,anderson2013,gorti2015,rosotti2017}. \citet{haworth2018a} introduce the concept of pre-computing photevaporation mass losses in terms of the surface density of the disks, an approach that we expand on in section \ref{photoevaporation}.  \citet{winter2019} model the environment of Cygnus OB2 and use the photoevaporation mass loss rate to constrain the timescale for gas expulsion in the young star forming region. \citet{nicholson2019} perform simulations of star forming regions where FUV photoevaporation is implemented in post-processing, and find a very short lifetime for the disks ($< \SI{2}{Myr}$) in moderate and low density regions ($\lesssim \SI{100}{\mathrm{M}_\odot \mathrm{\ pc}^{-3}}$).

Regarding dynamical effects, close encounters on a single N-body disk of test particles have been investigated in several studies \citep[e.g.][]{breslau2014,jilkova2016,bhandare2016,pfalzner2018}. \citet{winter2018,winter2018a} use a ring of test particles around a star to obtain linearized expressions of the effect that a stellar encounter can have on the mass and morphology of the disk, and then use them to simulate the disk using a smoothed particles hydrodynamics (SPH) code. A different approach for studying these effects is evolving the stellar dynamics of the cluster separately, and applying the effects of dynamical encounters afterwards \citep[e.g.][]{olczak2006,olczak2010,malmberg2011,steinhausen2014,vincke2015,vincke2016,vincke2018}. Directly adding SPH disks to a simulation of a massive star cluster is computationally expensive, since a high resolution is needed over long time scales. The closest effort corresponds to the work by \citet{rosotti2014}, in which individual SPH codes are coupled to half of the $\SI{1}{M_\odot}$ stars in a cluster with 100 stars. This allows them to study the effects of viscous spreading of the disks and dynamical truncations in a self-consistent way, but they are limited by the computational resources needed for this problem. Parametrized approaches have also been developed, where the cluster dynamics and effects of truncations \citep{portegieszwart2016} and viscous spreading \citep{concha-ramirez2019} are considered simultaneously. 

\citet{concha-ramirez2019} investigate the effect of viscous growth and dynamical truncations on the final sizes and masses of protoplanetary disks inside stars clusters using a parametrized model for the disks. They show that viscous evolution and dynamical encounters are unable to explain the compact disks observed in star forming regions. They argue that other processes must affect the early evolution of the disks. Here we expand the model by improving the description of the viscous disks and by adding external photoevaporation due to the presence of bright nearby stars.

We model the circumstellar disks using the Viscous Accretion Disk Evolution Resource (VADER) \citep{krumholz2015}. This code solves the equations of angular momentum and mass transport in a thin, axisymmetric disk. Dynamical truncations are parametrized, and the mass loss due to external photevaporation is calculated using the Far-ultraviolet Radiation Induced Evaporation of Discs (FRIED) grid \citep{haworth2018}. This grid consists of pre-calculated mass loss rates for disks of different sizes and masses, immersed in several different external FUV fields. We use the Astrophysical Multipurpose Software Environment \citep[AMUSE\footnote{\url{http://amusecode.github.io/}},][]{portegieszwart2018a} framework to couple these codes along with cluster dynamics and stellar evolution. All the code developed for the simulations, data analyses, and figures of this paper is available in Github\footnote{https://github.com/franciscaconcha/arXiv1907.03760}.

\section{Model}
\label{sec:model}

\subsection{Viscous growth of circumstellar disks}\label{viscous}
We implement circumstellar disks using the Viscous Accretion Disk Evolution Resource (VADER) by \citet{krumholz2015}. VADER is an open source viscous evolution code which uses finite-volume discretization to solve the equations of mass transport, angular momentum, and internal energy in a thin, axisymmetric disk. An AMUSE interface for VADER was developed in the context of this work and is available online.  

For the initial disk column density we use the standard disk profile introduced by \citet{lynden-bell1974}:

\begin{equation}\label{profile}
\Sigma(r, t=0) = \Sigma_0 \frac{r_c}{r} \exp\left(\frac{-r}{r_c}\right),
\end{equation}

\noindent
with

\begin{equation}
\Sigma_0 = \frac{m_d}{2 \pi {r_c}^2 \left(1 - \exp\left(-r_d/r_c\right)\right)},
\end{equation}

\noindent
where $r_c$ is the characteristic radius of the disk, $r_d$ and $m_d$ are the initial radius and mass of the disk, respectively, and $\Sigma_0$ is a normalization constant. Considering $r_c \approx r_d$ \citep{anderson2013}, the density profile of the disk is:

\begin{equation}
\Sigma(r, t=0) \approx \frac{m_d}{2 \pi r_d \left(1 - e^{-1}\right)} \frac{\exp(-r/r_d)}{r}.
\end{equation}

This expression allows the disk to expand freely at the outer boundary while keeping the condition of zero torque at the inner boundary $r_i$.

\begin{figure}
  \includegraphics[width=\linewidth]{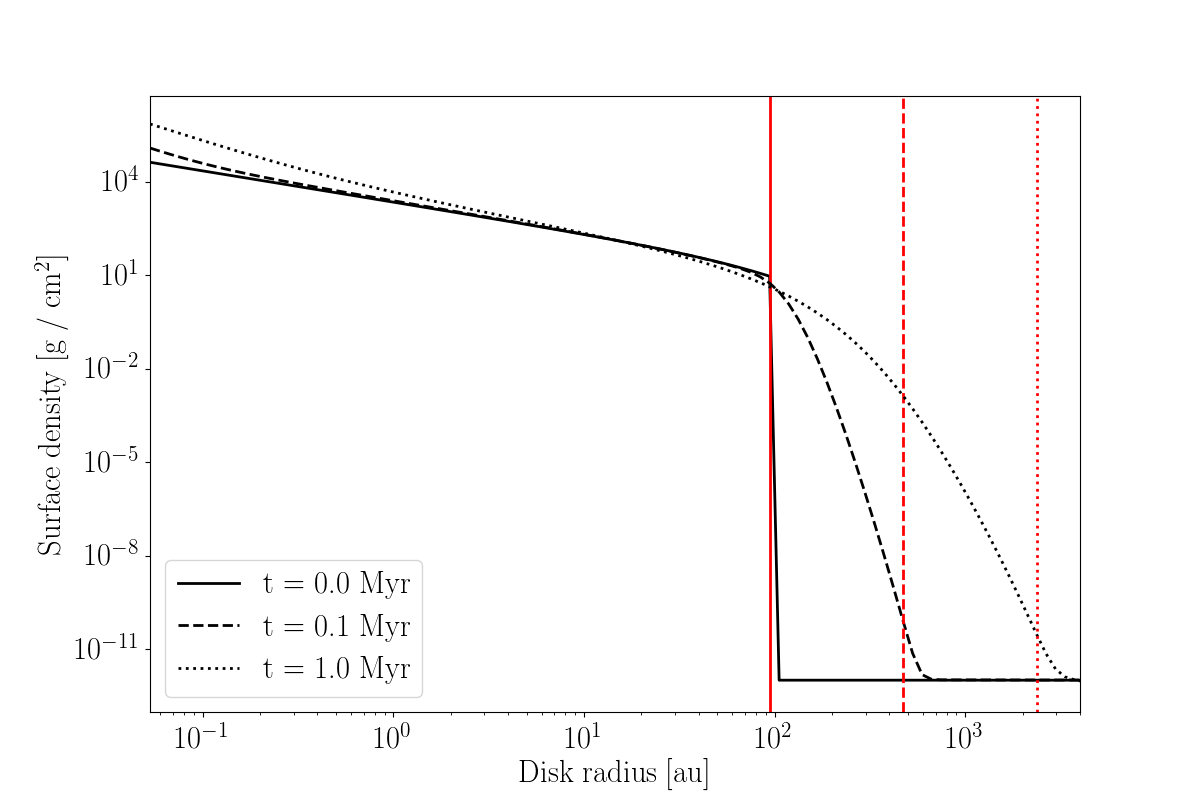}
  \caption{Definition of disk radius. The black lines show the disk profile and the corresponding red lines show the measured disk radius. Solid lines represent a disk initialized with $R = \SI{100}{au}$. The dashed lines show the same disk after $\SI{0.1}{Myr}$ of isolated evolution, and the dotted lines after $\SI{1.0}{Myr}$ of isolated evolution.}
  \label{fig:radius_def}
\end{figure}

To establish the radius of the disks we set the column density outside $r_d$ to a negligible value $\Sigma_{\mathrm{edge}} = \SI{e-12}{\gram\per\square\cm}$. The FRIED grid that we use to calculate the photoevaporation mass loss (see section \ref{grid}) is a function of disk radius and outer surface density. There is a numerical challenge in determining what the disk outer surface density and radius actually are, since there is a large gradient in it down to the $\Sigma_{\mathrm{edge}}$ value. Computing a mass loss rate for a very low outer surface density in this steep gradient would return an artificially low result. Considering this we define the disk radius as the position of the first cell next to $\Sigma_{\mathrm{edge}}$, as shown in Figure \ref{fig:radius_def}.

The temperature profile of the disks is given by

\begin{equation}
T(r) = T_m \left(\frac{r}{\SI{1}{au}}\right)^{-1/2},
\end{equation}

\noindent
where $T_m$ is the midplane temperature at $\SI{1}{au}$. Based on \citet{anderson2013} we adopt $T_m = \SI{300}{\K}$.

Each disk is composed of a grid of 50 logarithmically spaced cells, in a range between $0.5$ and $\SI{5000}{au}$. In Appendix \ref{appendix:resolution} we show that the resolution is enough for our calculations. The disks have a Keplerian rotation profile and turbulence parameter $\alpha=\num{5e-3}$. The fact that the outer radius of the grid is much larger than the disk sizes (which were initially around $\SI{100}{au}$, see section \ref{initdisks}) allows the disks to expand freely without reaching the boundaries of the grid. The mass flow through the outer boundary is set to zero in order to maintain the density $\Sigma_{\mathrm{edge}}$ needed to define the disk radius. The mass flow through the inner boundary is considered as accreted mass and added to the mass of the host star.

\subsection{Dynamical truncations}\label{truncations_dynamical}
A close encounter between disks induces a discontinuity in their evolution. To modify the disks we calculate parametrized truncation radii. For two stars of the same mass, \citet{rosotti2014} approximated the truncation radius to a third of the encounter distance. Together with the mass dependence from \citet{bhandare2016}, the dynamical truncation radius takes the form:

\begin{equation}\label{truncationradius}
{r^\prime} = \frac{r_{enc}}{3}\left(\frac{m_1}{m_2}\right)^{0.32},
\end{equation}

\noindent 
where $m_1$ and $m_2$ are the masses of the encountering stars.

To implement truncations we first calculate the corresponding truncation radius caused by the encounter, according to equation \ref{truncationradius}. We consider all the mass outside this radius to be stripped from the disk. To define $r^\prime$ as the new disk radius we change the column density of all the disk cells outside $r^\prime$ to the edge value $\Sigma_{\mathrm{edge}} = \SI{e-12}{\gram\per\square\cm}$.

\subsection{External photoevaporation}\label{photoevaporation}

A circumstellar disk can be evaporated by radiation coming from its host star or from a bright star in the vicinity. The heating of the disk surface can lead to the formation of gaps at different locations, which can cause the progression to a transition disk \citep[e.g.][]{clarke2001, gorti2009}. The radiation can also truncate the disk by removing material from the outer, loosely-bound regions \citep[e.g.][]{adams2004}.

{Models of internal photoevaporation have shown that most of the mass loss in this case occurs in the inner $\SI{20}{au}$ of the disk \citep{font2004, owen2010}. Radiation from the host star can evaporate material from the outskirts of the disk, however \citet{owen2010} show that more than 50\% of the total mass loss occurs in the $5 - \SI{20}{au}$ region. In \citet{font2004} around 90\% of the mass loss occurs within the inner $\SI{18}{au}$ of the disk. Given that the mass loss rate from the outer disk is typically comparable to or larger than that from the inner disk, we expect external photoevaporation to dominate in the outer regions. External photoevaporation has been shown to be more effective in evaporating the disks than radiation from the host star \citep[e.g.][]{johnstone1998, adams2001}. Truncation by external photoevaporation can also result in changes of the viscosity parameter $\alpha$, which further affects the viscous evolution of the disks \citep{rosotti2017}. In this work we ignore the effects of photoevaporation on internal disk structure, and deal exclusively with disk survival rates. Because of this and the discussion above we focus on external photoevaporation due to bright stars in the vicinity.

OB-type stars emit heating radiation in the form of extreme-ultraviolet (EUV), far-ultraviolet (FUV), and X-rays. In the case of external photoevaporation the dispersal of disk material is dominated by the FUV photons \citep{armitage2000, adams2004, gorti2009a}. The main part of our work deals with photoevaporation due to FUV photons; in addition we also incorporate the effect of EUV photons (see Eq. \ref{eq:euv}).

The amount of mass lost from the disks as a result of external photoevaporation depends on the luminosity of the bright stars in the cluster. This luminosity, together with the distance from each of the massive stars to the disks, is used to obtain the amount of radiation received by each disk. We can then calculate the amount of mass lost. Below we explain what we consider to be massive stars and how we calculate the mass loss rate.

\subsubsection{FUV luminosities}\label{FUVluminosities}
We follow \citet{adams2004} in defining the FUV band ranging from 6 eV up to 13.6 eV, or approximately from 912 $\angstrom$ to 2070 $\angstrom$. We calculate the FUV radiation of the stars in the simulations based on their spectral types. Given the presence of spectral lines in this band, we use synthetic stellar spectra rather than relying on black body approximations. The spectral library used is UVBLUE \citep{rodriguez-merino2005}, chosen for its high coverage of parameter space and high resolution, spanning the appropriate wavelength ranges. It spans a three dimensional parameter space of stellar temperature, metallicity, and surface gravity.

We use the UVBLUE spectral library to precompute a relation between stellar mass and FUV luminosity. We do this by considering all the stars in the cluster to have solar metallicity ($\mathrm{Z} = 0.02$). We then select the temperature and surface gravity spectra closer to the zero age main sequence value of each star, according to the parametrized stellar evolution code SeBa \citep{portegieszwart1996, toonen2012}. 
Given that the masses and radii of the stars are known, using the chosen spectra we build a relationship between stellar mass and FUV luminosity. This relation takes the shape of a segmented power law, as is shown in Figure \ref{fig:fit}. A similar fit was obtained by \citet{parravano2003}. In runtime the stars are subject to stellar evolution and the FUV luminosity for each star was calculated directly from this fit using the stellar mass. 

\begin{figure}
  \includegraphics[width=\linewidth]{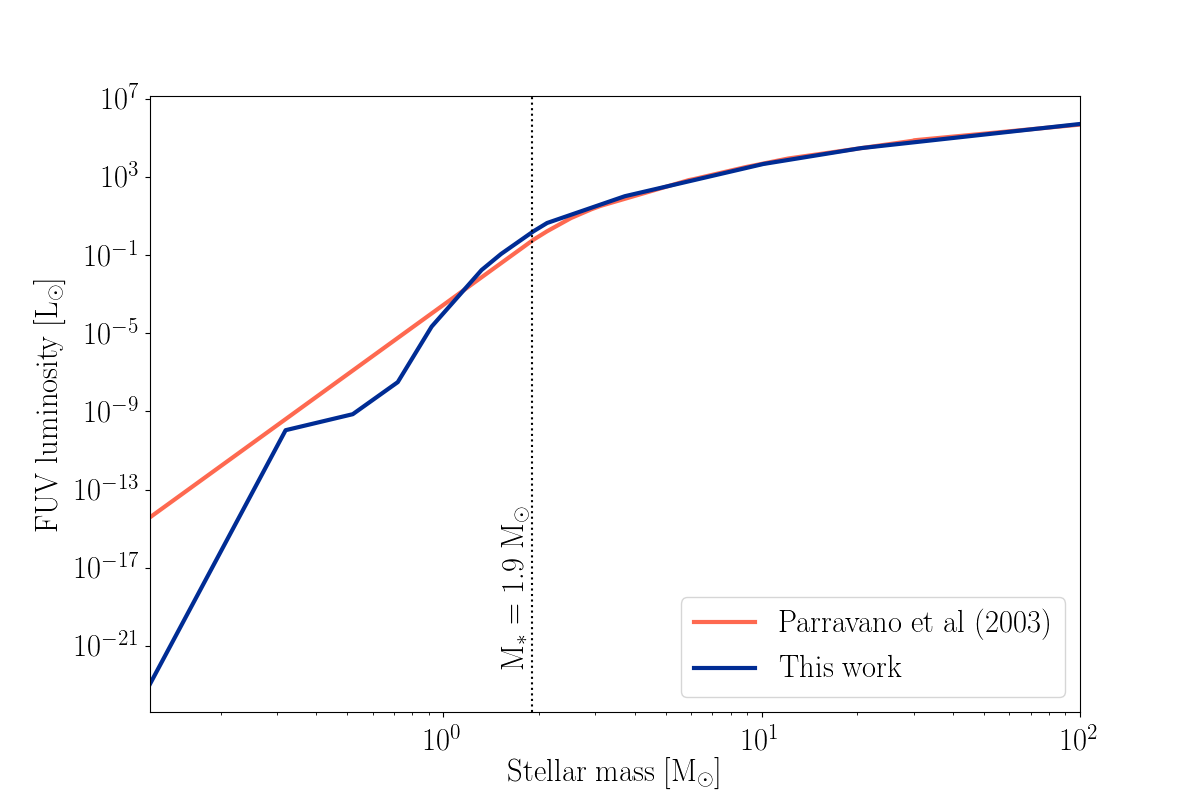}
  \caption{FUV luminosity vs. stellar mass fit calculated from the ZAMS spectra. $\mathrm{M}_* = 1.9 \mathrm{\ M}_\odot$ is the lower mass limit of the FRIED grid, and the lower mass limit for the stars to be considered emitting FUV radiation in our simulations (see Section \ref{grid} for details).}
  \label{fig:fit}
\end{figure}

The mass range of the fit in Figure \ref{fig:fit} is $0.12 - 100 \mathrm{\ M}_{\odot}$, however, we are only interested in the range $1.9 - 50 \mathrm{\ M}_{\odot}$. As is further explained in section \ref{grid}, we only consider stars with masses higher that $1.9 \mathrm{\ M}_{\odot}$ to be emitting FUV radiation, and $50 \mathrm{\ M}_{\odot}$ is the theoretical upper limit for the stellar mass distribution. Stars with masses $\leq 1.9 \mathrm{\ M}_{\odot}$ are given disks and are affected by the massive stars. The massive stars are also subject to stellar evolution, which is implemented with SeBa through the AMUSE interface. The FUV luminosity of each massive star is calculated in every time step, from the precomputed fit, after evolving the stellar evolution code. The low mass stars are not subject to stellar evolution.

\subsubsection{Mass loss rate due to external photoevaporation}\label{grid}
To calculate the mass loss due to FUV external photoevaporation we use the Far-ultraviolet Radiation Induced Evaporation of Discs (FRIED) grid developed by \citet{haworth2018}. The FRIED grid is an open access grid of calculations of externally evaporating circumstellar disks. It spans a five dimension parameter space consisting of disk sizes ($1 - 400 \mathrm{\ au}$), disk masses ($~10^{-4} - 10^{2} \mathrm{\ M}_{Jup}$), FUV fields ($10 - 10^{4} \mathrm{\ G}_0$), stellar masses ($0.05 - 1.9 \mathrm{\ M}_{\odot}$) and disk outer surface densities. The seemingly three dimensional grid subspace of disk mass, edge surface density, and disk radius is in fact two dimensional, as any combination of disk radius and disk mass has only one edge density associated with it. Because of this, we only take into account a four dimensional grid of radiation field strength, host star mass, disk mass, and disk radius.

Following the stellar mass ranges of the FRIED grid we separate the stars in the simulations into two subgroups: \textit{massive stars} and \textit{low mass stars}. Massive stars are all stars with initial masses higher than $1.9 \mathrm{\ M}_{\odot}$, while low mass stars have masses $\leq 1.9 \mathrm{\ M}_{\odot}$. Only the low mass stars have circumstellar disks in the simulations. The massive stars are considered as only generating FUV radiation and affecting the low mass stars. In this way we make sure that we stay within the stellar mass limits of the FRIED grid. Low mass stars ($\lesssim 1 \mathrm{\ M}_\odot$) have a negligible UV flux \citep{adams2006}, so this approximation holds well for our purposes. Calculation of the FUV radiation emitted by the massive stars is further explained in section \ref{FUVluminosities}. These star subgroups are considered only for the calculation of FUV radiation and photoevaporation mass loss. In the gravity evolution code the two subgroups are undistinguishable. 

The FRIED grid allows to take a circumstellar disk with a specific mass, size, and density, around a star with a certain mass, and from the FUV radiation that it receives, obtain the photoevaporation mass loss. However, the parameters of the simulated disks do not always exactly match the ones in the grid. We perform interpolations over the grid to calculate the mass losses of the disks in the simulations. Because of computational constraints, we perform the interpolations on a subspace of the grid, such that it contains at least one data point above and below the phase space point of the disk in each dimension. The high resolution of the FRIED grid ensures that this interpolation is performed over an already smooth region.

When a massive star approaches a circumstellar disk, external photoevaporation is dominated by EUV radiation. Following \citet{johnstone1998} we define a distance limit below which EUV photons dominate:

\begin{equation}\label{eq:dmin}
    d_{min} \simeq 5 \times 10^{17} \left(\frac{\epsilon^2}{f_r \Phi_{49}}\right)^{-1/2} r_{d_{14}}^{1/2} \mathrm{\ cm}
\end{equation}

\noindent
where $\left(\frac{\epsilon^2}{f_r \Phi_{49}}\right)^{1/2} \approx 4$, $r_{d_{14}} = \frac{r_d}{10^{14} cm}$ with $r_d$ the disk radius, and $5 \times 10^{17} \mathrm{\ cm} \sim 3 \times 10^4 \mathrm{\ au} \sim 0.16 \mathrm{\ pc}$. When a star with a disk is at a distance $d < d_{min}$ from a massive star we calculate the mass loss using equation (20) from \citet{johnstone1998}:

\begin{equation}\label{eq:euv}
    \dot{M}_{EUV} = 2.0 \times 10^{-9} \frac{(1 + x)^2}{x} \epsilon r_{d_{14}} M_{\odot} \mathrm{\ yr}^{-1}
\end{equation}

\noindent
with $x \approx 1.5$ and $\epsilon \approx 3$. During most of their evolution, however, the circumstellar disks in the simulations experience photoevaporation only due to FUV photons. Since we do not consider interstellar gas and dust in the clusters, we do not account for extinction in the calculation of the radiation received by each small star.

\subsubsection{Disk truncation due to photoevaporation}\label{truncations_photoevaporation}
Once the mass loss due to photoevaporation is calculated for every disk, the disks are truncated at a point that coincides with the amount of mass lost in the process. We take the approach of \citet{clarke2007} and remove mass from the outer edge of the disk. We do this by moving outside-in from the disk edge and removing mass from each cell by turning its column density to the edge value $\Sigma_{\mathrm{edge}} = 10^{-12}$ defined in section \ref{viscous}. We stop at the point where the mass removed from the disk is equal to the calculated mass loss. 

We consider a disk to be completely evaporated when it has lost 99\% of its initial mass \citep{anderson2013} or when its mean column density is lower than $\num{1} \mathrm{\ g\ cm}^{-2}$ \citep{pascucci2016}. From this point forward the star continues its dynamical evolution normally, but is no longer affected by massive stars.

\subsubsection{Summary of cluster evolution}\label{evolution_summary}
The code runs in major time steps, which represent the time scale on which the various processes are coupled. Within each of these macroscopic time steps ($\SI{1000}{yr}$), internal evolutionary processes such as stellar evolution and gravitational dynamics proceed in their own internal time steps ($\SI{500}{yr}$ and $\SI{1000}{yr}$ respectively). Throughout each macroscopic time step, we perform the following operations:

\begin{enumerate}[leftmargin=*]
\item[1.] Gravitational dynamics code is evolved.
\item[2.] We check the stars for dynamical encounters:
\begin{enumerate}[leftmargin=*]
    \item[2.a] If a dynamical encounter occurs, we determine the truncation radius for each disk.
    \item[2.b] We update the radius and mass of the disks.
\end{enumerate}
\item[3.] Stellar evolution code is evolved.
\item[4.] Photoevaporation process is carried out as follows. For each massive star:
\begin{enumerate}[leftmargin=*]
    \item[4.a] We calculate the distance $d$ from the massive star to each low mass star.
    \item[4.b] If $d < d_{min}$ (see Eq. \ref{eq:euv}) we calculate the mass loss $\dot{M}_{EUV}$.
    \item[4.c] If $d \geq d_{min}$ the massive star's FUV luminosity $L_{FUV}$ is calculated (see section \ref{FUVluminosities}).
    \item[4.d] Using $d$ and $L_{FUV}$ we calculate $l_{FUV}$, the amount of FUV radiation received by the low mass star.
    \item[4.e] Using $l_{FUV}$ together with the low mass star's mass, disk mass, and disk radius, we build a subgrid of the FRIED grid and interpolate over it to calculate the mass loss $\dot{M}_{FUV}$.
    \item[4.f] The total mass loss in the time step is calculated using $\dot{M}_{EUV}$ and $\dot{M}_{FUV}$.
    \item[4.g] The mass is removed from the disk by moving outside-in and removing mass from the cells.
    \item[4.h] The disk mass and radius are updated.
\end{enumerate}
\item[5.] Disks are checked for dispersal. If a disk has been dispersed (see section \ref{truncations_photoevaporation}) the code for the disk is stopped and removed and the star continues evolving only as part of the gravitational dynamics code.
\item[6.] Simulation runs until $\SI{5}{Myr}$ of evolution of until all the disks are dispersed, whichever happens first.
\end{enumerate}

We present a scheme of this process and of the photoevaporation in Figures \ref{fig:diagram1} and \ref{fig:diagram2} respectively.

\begin{figure}
  \includegraphics[width=\linewidth]{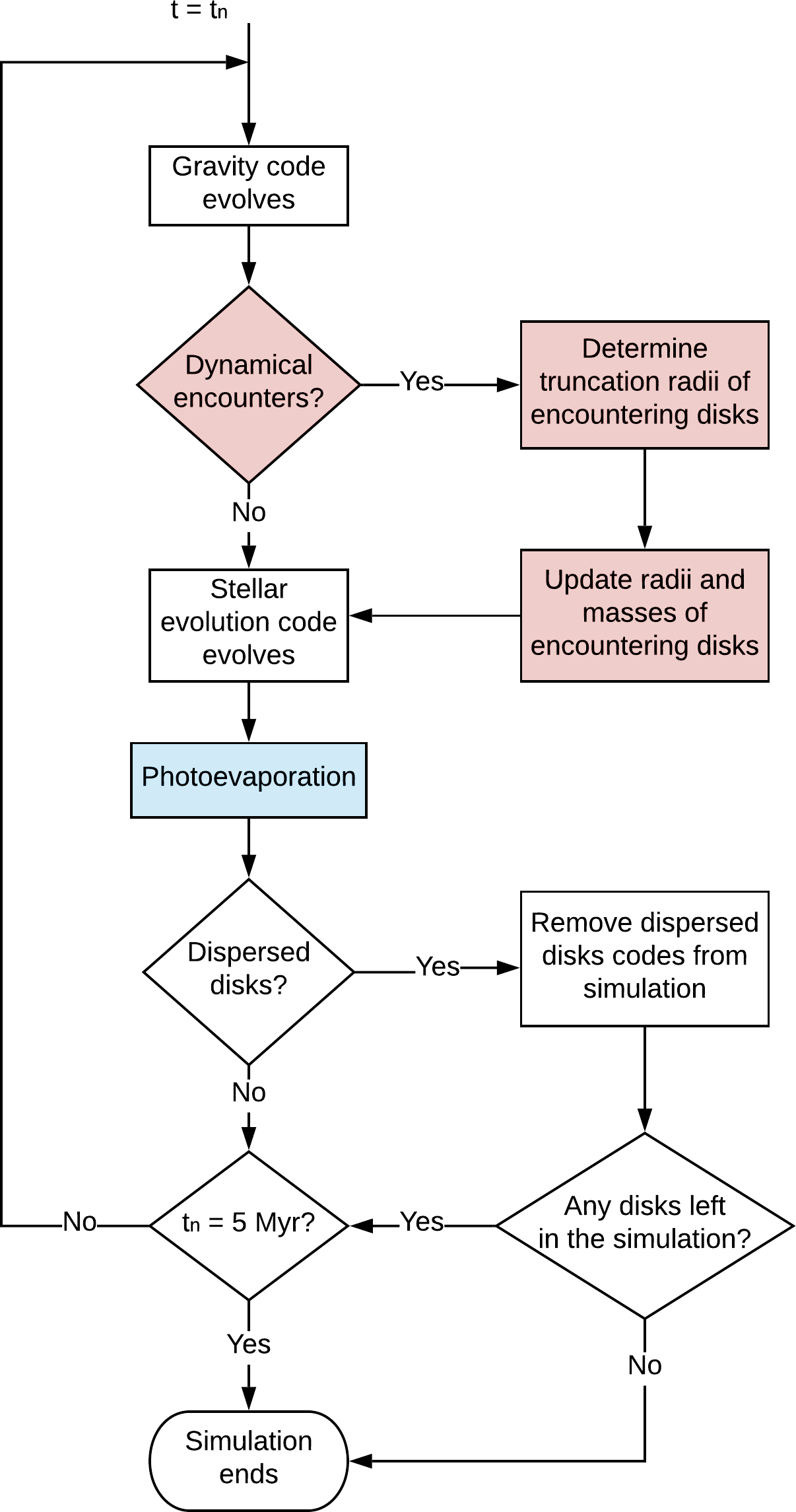}
  \caption{Operations performed in each macroscopic time step. Within each macroscopic time step $\mathrm{t}_{\mathrm{n}}$, internal evolutionary processes such as stellar evolution and gravitational dynamics proceed in their own internal time steps.}
  \label{fig:diagram1}
\end{figure}

\begin{figure}
  \includegraphics[width=\linewidth]{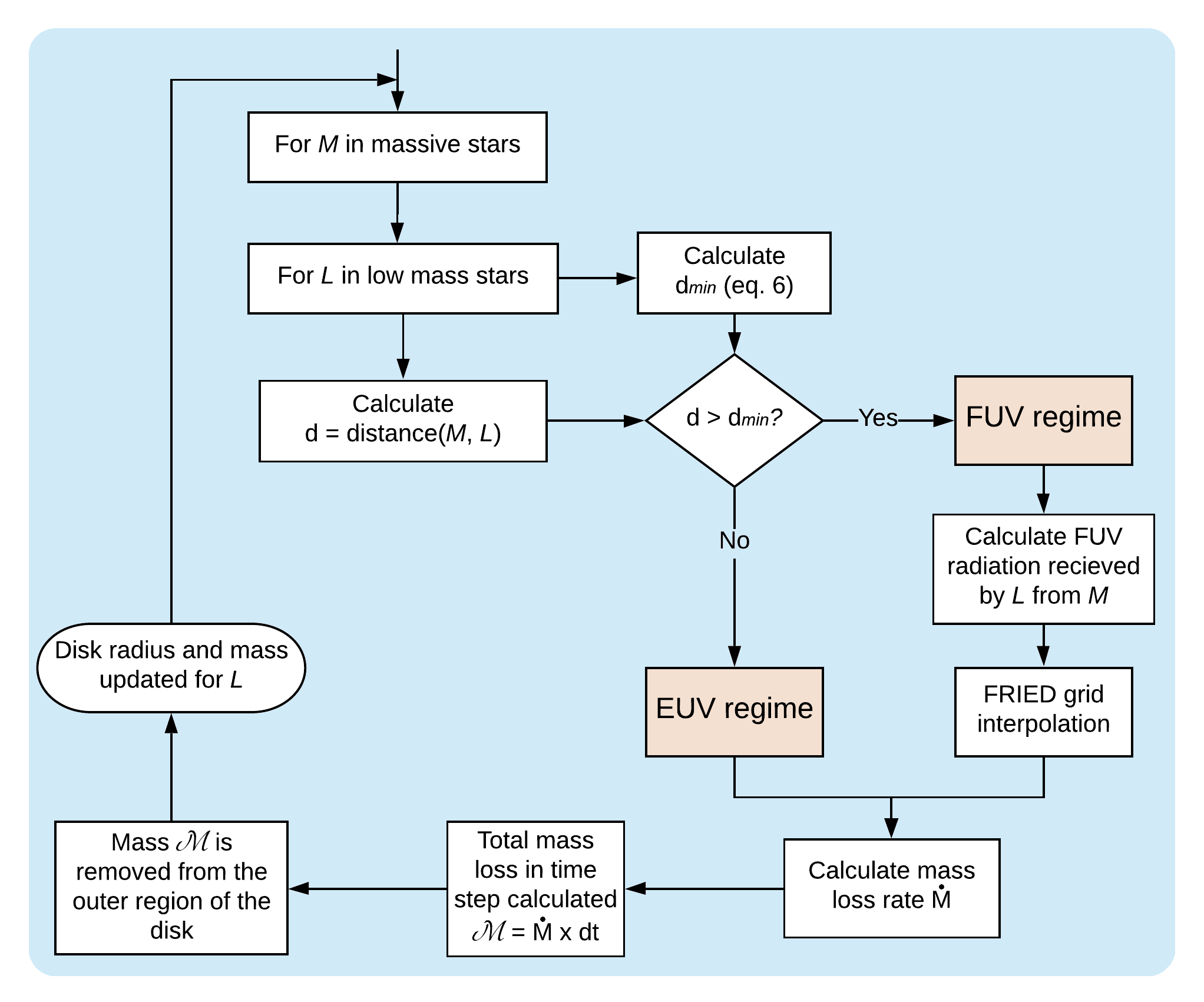}
  \caption{Flowchart of the photoevaporation process.}
  \label{fig:diagram2}
\end{figure}

\subsection{Initial conditions}\label{ics}
\subsubsection{Disks}\label{initdisks}
The initial radii of the circumstellar disks are given by:

\begin{equation}
R_d(t=0) = R'\left(\frac{M_*}{M_\odot}\right)^{0.5}
\end{equation}

\noindent
where $R'$ is a constant. The youngest circumstellar disks observed to date have diameters that range from $\sim\SI{30}{au}$ \citep[e.g.][]{lee2018} to $\sim120-\SI{180}{au}$ \citep[e.g.][]{murillo2013,vanthoff2018}. Based on this we choose $R' = \SI{100}{au}$, which for our mass range $0.05-1.9 \mathrm{\ M}_\odot$ for stars with disks results in initial disk radii between $\sim\SI{22}{au}$ and $\sim\SI{137}{au}$.

The initial masses of the disks are defined as:

\begin{equation}
\mathrm{M}_d(t=0) = 0.1 \mathrm{M}_*    
\end{equation}

\subsubsection{Cluster}
We perform simulations of young star clusters with stellar densities \rHSM and \rFSM using Plummer sphere spatial distributions \citep{plummer1911}. We will refer to these distributions as \HSM and \FSM, respectively. 
All the regions are in virial equilibrium (viral ratio $Q=0.5$). 

Stellar masses are randomly drawn from a Kroupa mass distribution \citep{kroupa2001} with maximum mass $50 \mathrm{\ M}_\odot$. The mean mass of the distribution is $\overline{\mathrm{M}}_* \approx 0.3 \mathrm{\ M}_\odot$. In Table \ref{table:stars} we show the details of the stellar masses in each simulation. 

\begin{table*}
\begin{tabular}{@{}llllll@{}}
\toprule
Region               & Sim. \# & Low mass stars & $\overline{\mathrm{M}}_{\mathrm{low\ mass}} [\mathrm{\ M}_\odot]$ & Massive stars & $\overline{\mathrm{M}}_{\mathrm{massive}} [\mathrm{\ M}_\odot]$ \\ \midrule
\multirow{3}{*}{\HSM} & 1      & 97\%              & 1.18 $\pm$ 0.17             & 3\%                & 4.08 $\pm$ 2.69              \\  
                     & 2      & 99\%              & 0.24 $\pm$ 0.30             & 1\%                & 3.43                      \\  
                     & 3      & 96\%              & 0.24 $\pm$ 0.31             & 4\%                & 4.51$\pm$ 2.59               \\ \midrule
\multirow{3}{*}{\FSM}  & 1      & 94\%              & 0.27 $\pm$ 0.33             & 6\%                & 5.94 $\pm$ 5.65              \\ 
                     & 2      & 96\%              & 0.25 $\pm$ 0.31             & 4\%                & 3.51 $\pm$ 0.26            \\  
                     & 3      & 99\%              & 0.24 $\pm$ 0.27             & 1\%                & 4.90                      \\ \bottomrule
\end{tabular}\caption{Stellar mass properties in each simulation run.}\label{table:stars}
\end{table*}

The simulations end at $\SI{5}{Myr}$ or when all the disks are dispersed, whichever happens first.

\begin{figure*}
  \includegraphics[width=\linewidth]{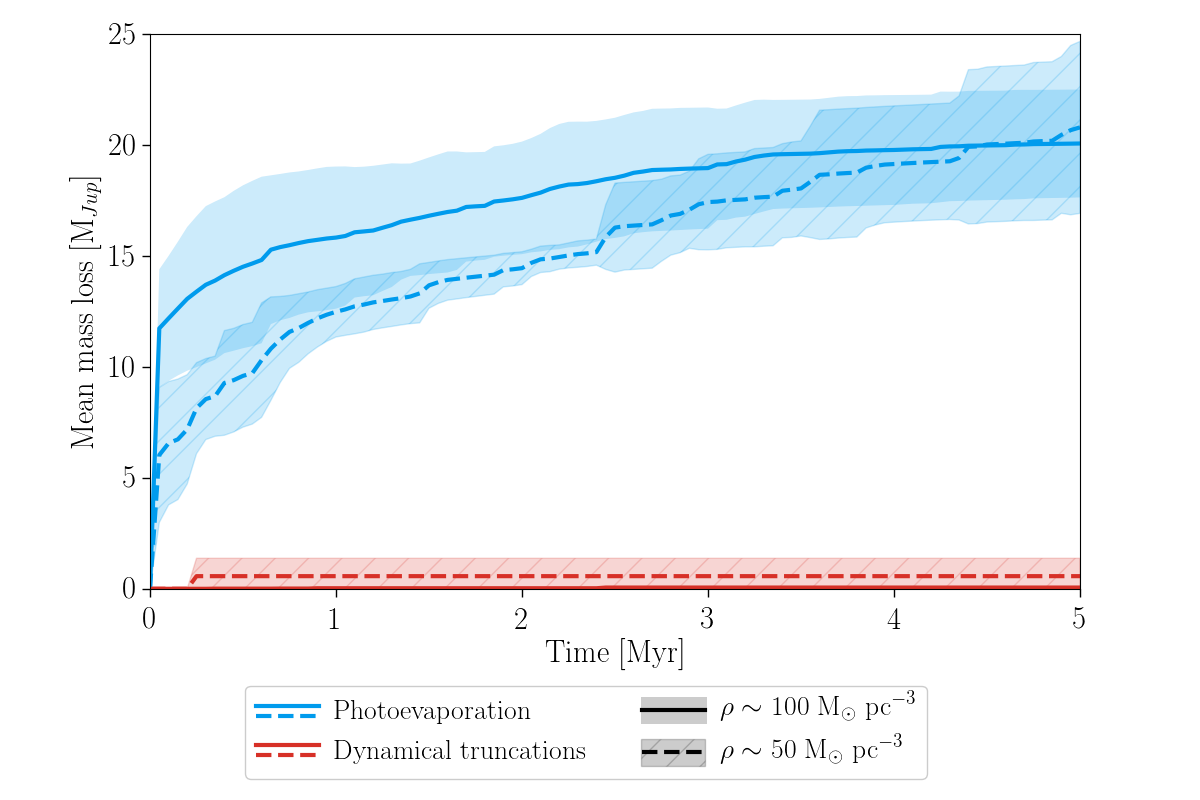}
  \caption{Mean mass loss in time due to external photoevaporation (blue) and dynamical truncations (red). The solid and dashed lines correspond to the \HSM and \FSM regions, respectively. The shaded areas indicate the standard deviation of the simulations.}
  \label{fig:massloss}
\end{figure*}

\section{Results}
\label{sec:results}

\subsection{Disk mass loss in time}
As a way to quantify the mass loss effect of each of the processes included in the simulations, we measure the mass loss due to external photoevaporation and dynamical truncations separately. In Figure \ref{fig:massloss} we show the mass loss over time for external photoevaporation and dynamical truncations. The solid and dashed lines correspond to the mean mass loss among all stars in the \HSM and \FSM regions, respectively. The shaded areas show the extent of the results in the different simulation runs.

The mass lost from the circumstellar disks is dominated by external photoevaporation over the entire lifetime of the simulated clusters. Dynamical truncations only have a local effect on truncating disk radii and masses, whereas photoevaporation is a global effect influencing all disks in the cluster.

The amount of FUV radiation received by each disk and the ensuing mass loss are variable. The effect depends on the proximity to massive stars, which changes with time as the stars orbit in the cluster potential. For the \HSM region the average FUV radiation over \SI{5}{Myr} of evolution was $\sim475$ \Go, with a minimum of 10 \Go and a maximum of $\sim 10^4$ \Go. For the \FSM region the average over \SI{5}{Myr} was $\sim56$ \Go, minimum $\sim2$ \Go and maximum 267 \Go. These values are only shown as an indicative of the environment that the simulation disks were dispersed in, however a short exposure to a strong FUV field can be instantly more destructive than a sustained low FUV field. The FUV field can also vary in time due to processes intrinsic to star formation, such as a massive star being embedded during its early stages \citep[e.g.][]{ali2019}. In our simulations, however, photoevaporation is driven by the background FUV field. In Figure \ref{fig:FUVfield} we show the cumulative distributions of disks that lose more than $5 \mathrm{\ M_{Jup}}$ in time. It can be seen that large mass losses are not driven by close encounters with bright stars but by the environmental FUV radiation. From Figure \ref{fig:FUVfield} it can be seen that before $\SI{2}{Myr}$ of evolution the disks in \HSM lose mass more strongly than the ones in \FSM. However, starting around $\SI{2}{Myr}$ and until the end of the simulations there is large scatter in the mass loss behaviour for each region. This is related to the dynamics of each cluster. For \HSM the crossing time is $t_\mathrm{cross} = 1.20 \pm 0.04 \mathrm{\ Myr}$, and for \FSM the crossing time is $t_\mathrm{cross} = 0.98 \pm 0.09 \mathrm{\ Myr}$. This results in the fact that, after one crossing time, all the stars in the clusters have been affected by the radiating stars similarly, causing the scatter in the mass loss effects. While the density of each cluster defines the effects of photoevaporation in the early evolutionary stages, after one crossing time the initial density of the region is not as important and photoevaporation works uniformly.

In Figure \ref{fig:maxmdot} we show the time step of maximum mass loss for each disk. It can be seen that, other than the effect of switching on the simulation at the beginning (see section \ref{disc:ic}), there is not a particular time at which a disk loses much mass. In Figure \ref{fig:cdf_mass} we show how the cumulative distributions of disk masses at different moments in the simulation. The solid lines correspond to \HSM the dashed lines to \FSM. Each line corresponds to the total disks in all simulations. It can be seen that disk mass distributions decrease monotonically. 

It is important to note that the FRIED grid has a lower limit of 10 \Go for the FUV field, which is higher than the minimum experienced in the \FSM region. However, more than 90\% of the stars in these simulations are within the limits of the grid at all times. In the few cases where stars were outside the limits of the grid, the mass loss obtained reflects a lower bound defined by the grid, but this does not affect our results.

\begin{figure}
  \includegraphics[width=\linewidth]{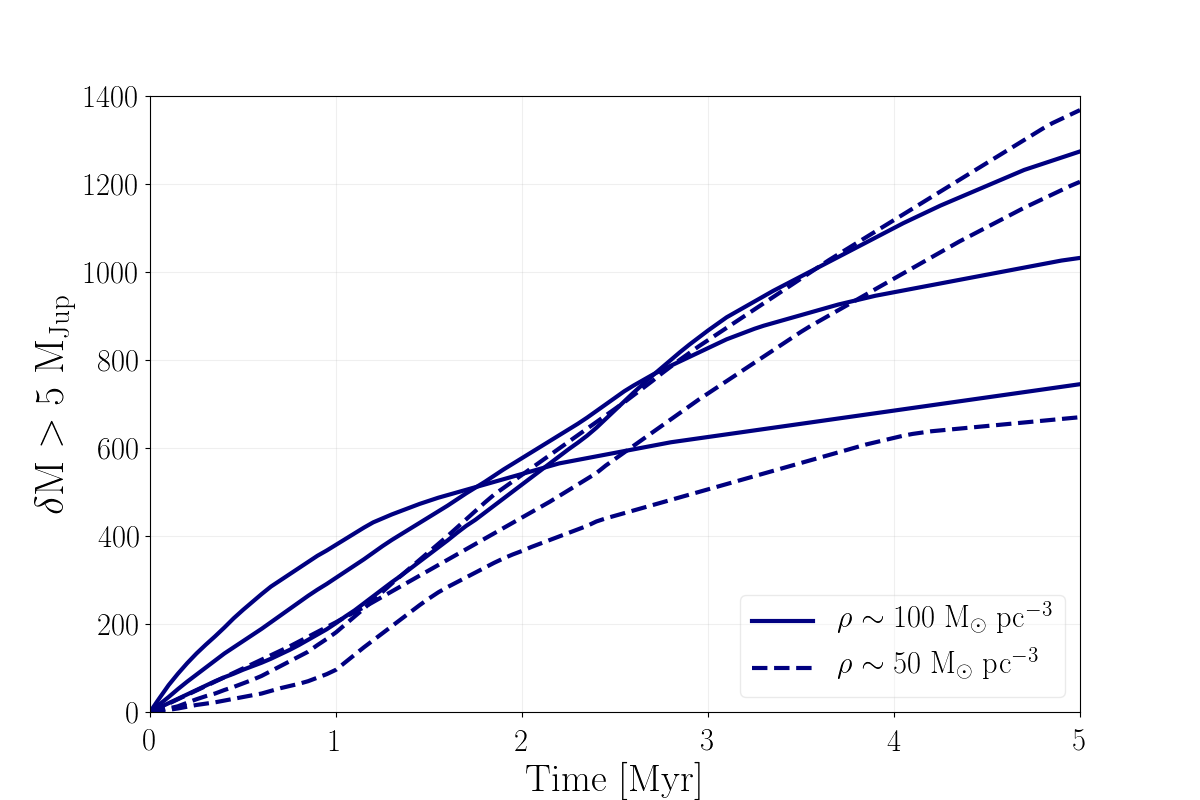}
  \caption{Cumulative distribution of disks that lose more than $5 \mathrm{\ M_{Jup}}$ in time in each simulation. The solid lines correspond to the \HSM region simulations and the dashed lines to the \FSM region simulations.}
  \label{fig:FUVfield}
\end{figure}

\begin{figure}
  \includegraphics[width=\linewidth]{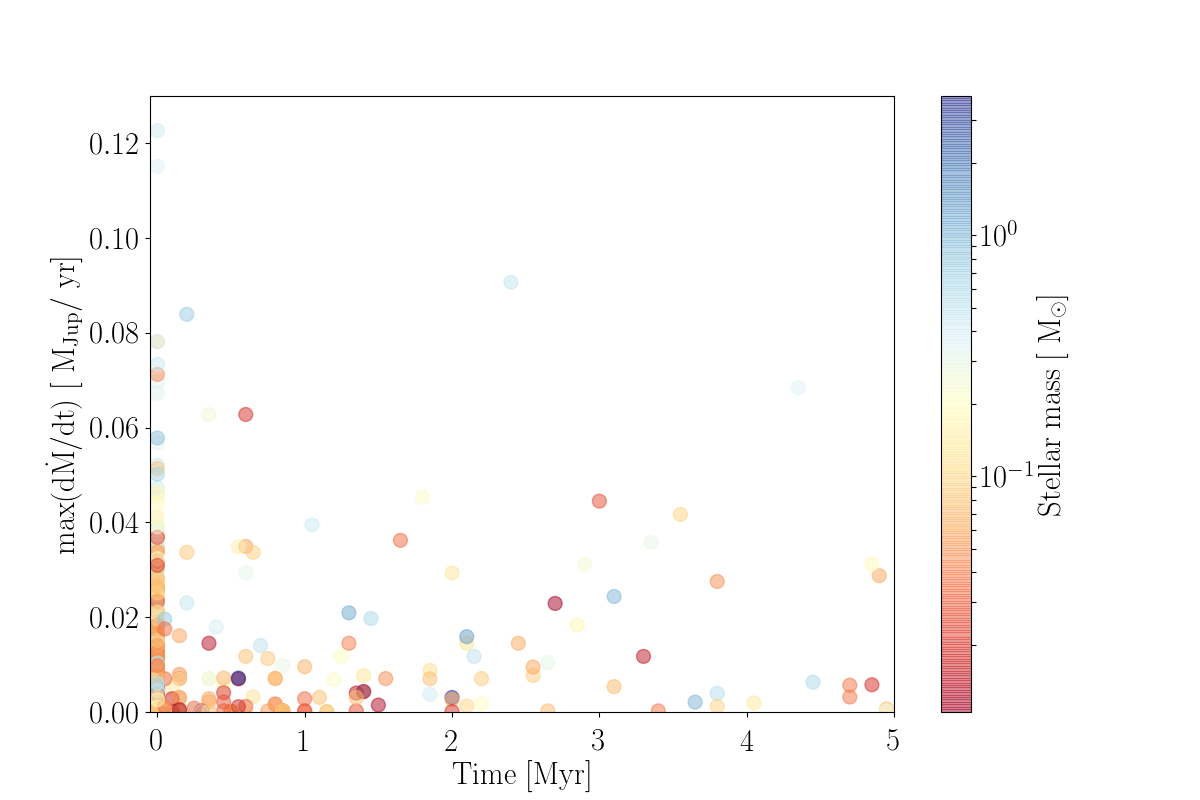}
  \caption{Maximum mass loss per time step for every disk. Each point corresponds to one disk in a simulation. The position of each point in time corresponds to its moment of maximum mass loss. This moment in time is not necessarily when the disk was dispersed.}
  \label{fig:maxmdot}
\end{figure}

\begin{figure}
  \includegraphics[width=\linewidth]{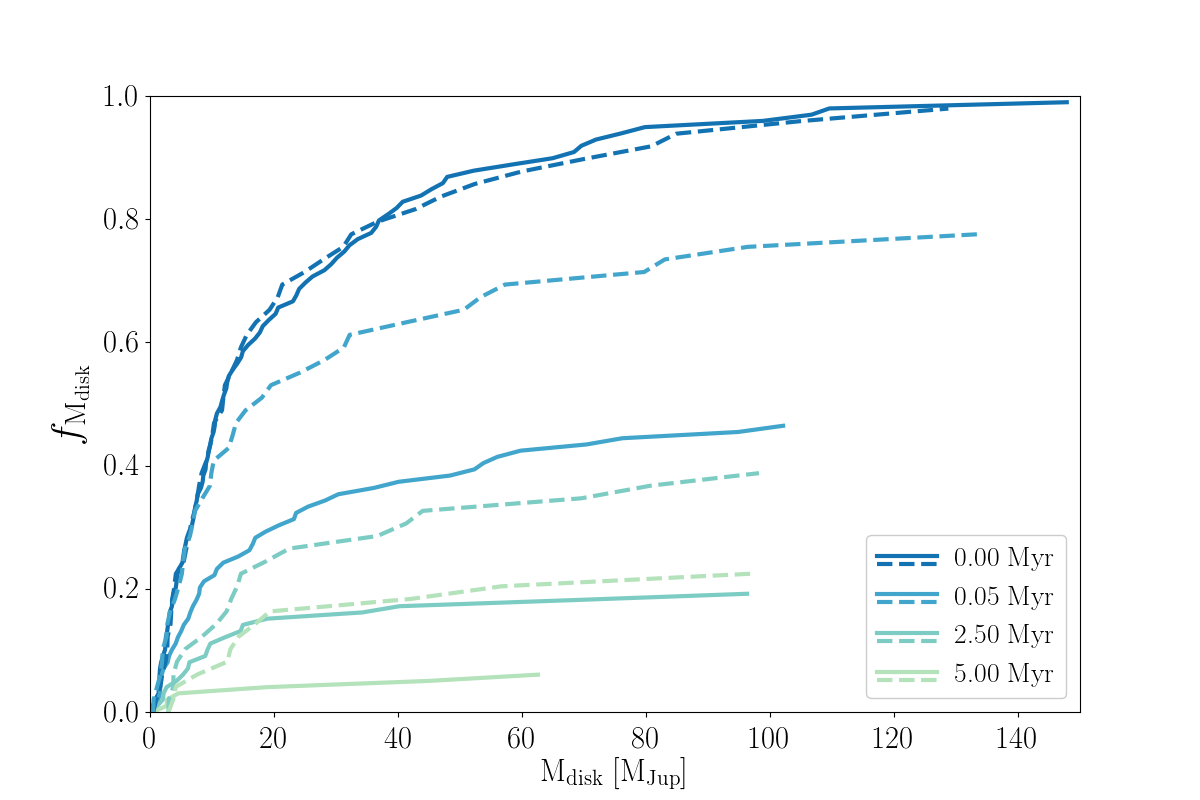}
  \caption{Cumulative distribution of disk masses at different simulation times. The solid lines correspond to the \HSM region and the dashed lines correspond to the \FSM region. Each curve shows the total number of disks in all simulations.}
  \label{fig:cdf_mass}
\end{figure}

Photoevaporation mass loss can have different effects over the gas and dust components of a circumstellar disk. We expand on the consequences of this for our results in section \ref{caveats}.

\subsection{Disk lifetimes}
The lifetime of circumstellar disks in young cluster regions is an important criterion to determine how photoevaporation affects planet formation. In Figure \ref{fig:diskfractions} we show disk fractions at different times of cluster evolution (black lines), together with observed disk fractions from \citet{ribas2014} and \citet{richert2018}. The orange line shows the mean of the observations, calculated using a moving bin spanning 10 observation points. The calculation of the mean starts by binning the first 10 points, and then sliding horizontally through the observations one point at a time such that 10 points are always considered.

The relaxation time is defined as:

\begin{equation}
    \mathrm{t}_\mathrm{relax} = 0.138 \frac{N}{\log(\gamma N)} \mathrm{t}_\mathrm{dyn}
\end{equation}

\noindent
\citep{spitzer1987}, where $N$ is the number of stars, $\gamma = 0.4$, and the dynamical time is:

\begin{equation}
    \mathrm{t}_\mathrm{dyn} = \sqrt{\frac{R^3}{GM}},
\end{equation}

\noindent
{where $R$ is the radius of the cluster and $M$ is its total mass. The relaxation time depends on the number of stars and the radius and mass of the stellar cluster. These are values that change through the dynamical evolution of a cluster, meaning that the relaxation time can grow and shrink at different time steps. These variations result in the jagged lines in Figure \ref{fig:diskfractions}.

For the simulations shown in Figure \ref{fig:diskfractions}, $\mathrm{t}/\mathrm{t}_{\mathrm{relax}} = 0.5$ is reached at $\mathrm{t} = \SI{2.01 \pm 0.37}{Myr}$ of evolution for \HSM and at $\mathrm{t} = \SI{2.05 \pm 0.35}{Myr}$ for \FSM. Disk fractions in the \HSM simulations drop to around $20\%$ before $\SI{2}{Myr}$ of cluster evolution. In the regions with \FSM the disks survive longer, but still most of the disks have disappeared by the end of the simulations. 

\begin{figure*}
  \includegraphics[width=\linewidth]{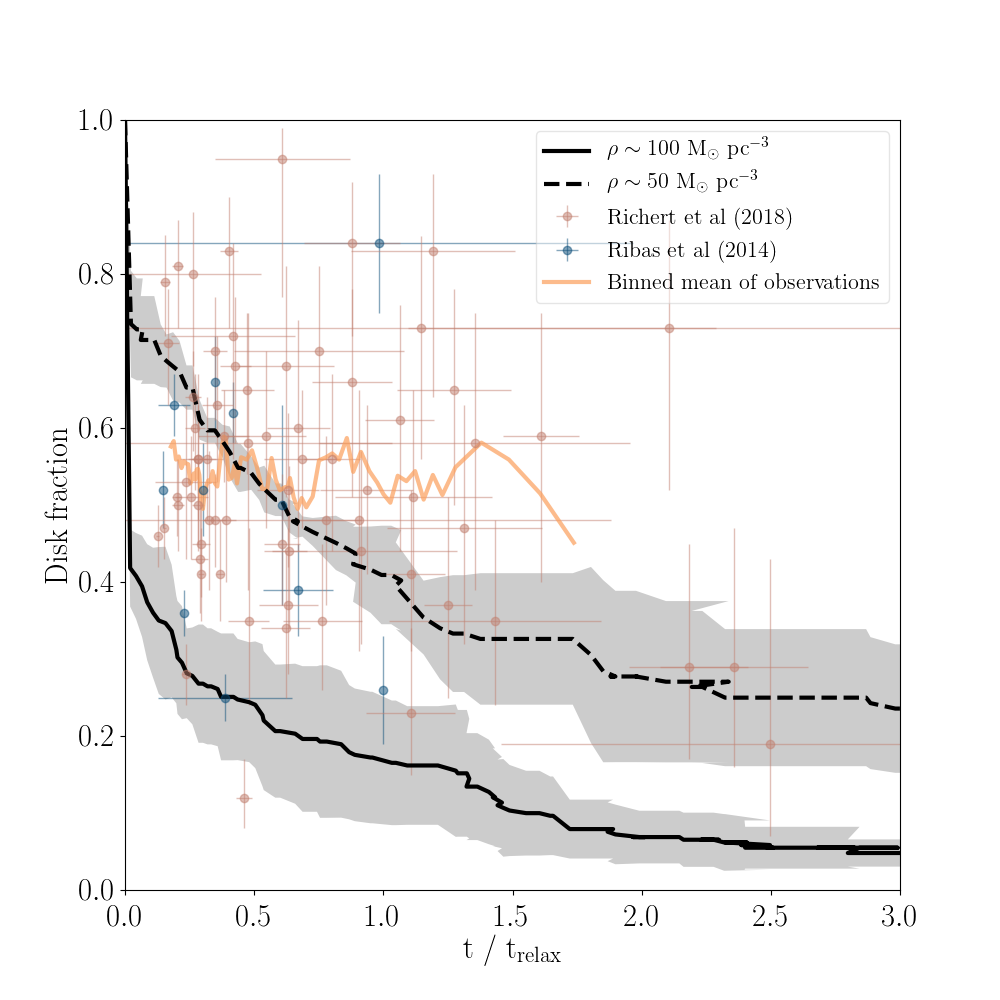}
  \caption{Fraction of stars with disks as a function of time. The solid black line shows the results for \HSM and the dashed black line for \FSM. The grey shaded areas indicate the standard deviation of the simulations. Observed disk fractions in star forming regions of different ages are shown for comparison. The orange line shows the mean of the observations, calculated using a moving bin spanning 10 observation points.}
  \label{fig:diskfractions}
\end{figure*}

Planet formation could still occur in disks that have been affected by photoevaporation as long as they are not completely dispersed. For gas giant cores and rocky planets to form, protoplanetary disks need to have a reservoir of dust mass $\mathrm{M}_{dust} \gtrsim 10 \mathrm{M}_{\oplus}$ \citep{ansdell2018}. In Figure \ref{fig:masslimit} we show the fraction of disks with solid masses $\mathrm{M}_{disk} > 10 \mathrm{M}_{\oplus}$ in time, for both simulated stellar densities. We use a 1:100 dust:gas mass ratio to turn the total disk mass into dust mass. For the \HSM regions the number of disks that fulfill this mass requirement drops to around $20\%$ at $\SI{1}{Myr}$, with less than $10\%$ of disks of said mass still present at the end of the simulations. For the less dense regions, at the end of the simulations around $20\%$ of disks with masses $\mathrm{M}_{disk} > 10 \mathrm{M}_{\oplus}$ survive. 

\begin{figure}
  \includegraphics[width=\linewidth]{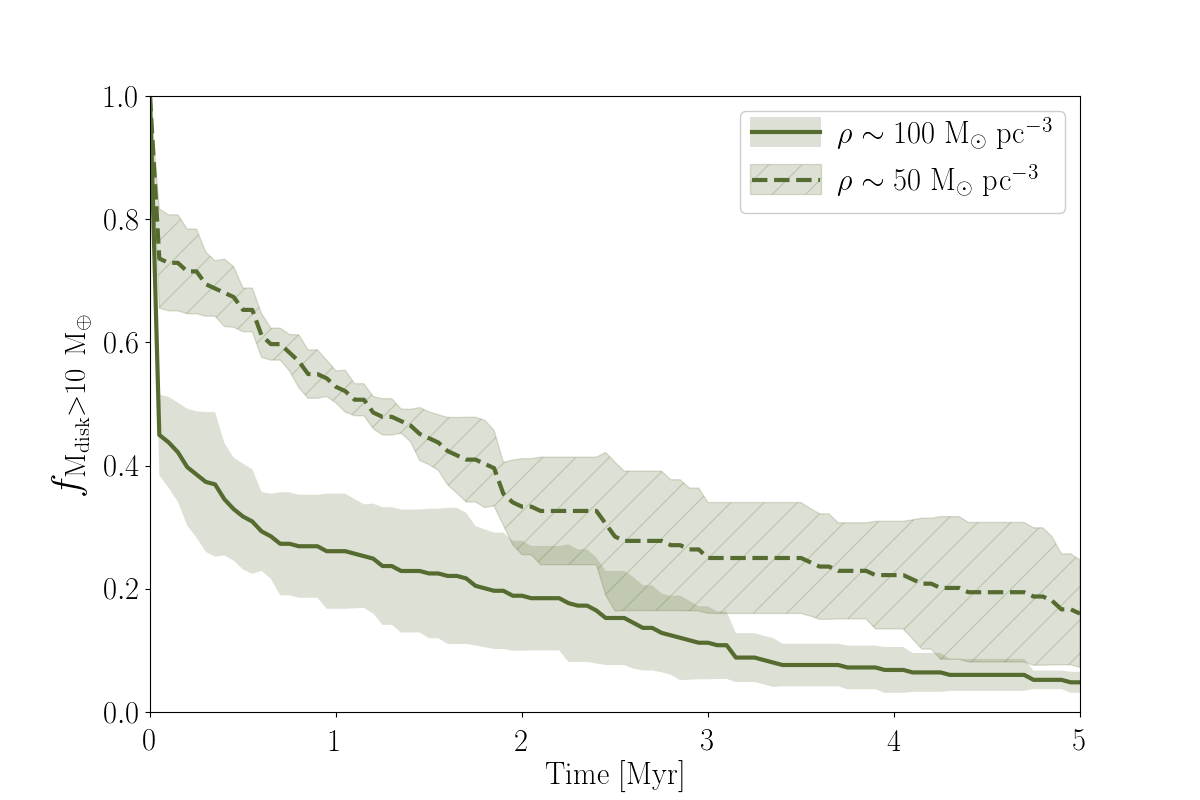}
  \caption{Fraction of disks with masses $\mathrm{M}_{disk} > 10 \mathrm{M}_\oplus$ in time, for regions of different stellar densities. The shaded areas indicate the standard deviation of the simulations.}
  \label{fig:masslimit}
\end{figure}

In order to make a parallel with the Solar System, in Figure \ref{fig:sizelimit} we show the number of disks in time with radii higher than $\SI{50}{au}$, for both density regions. The drop in disk sizes is slower than the drop in disk masses as seen in Figure \ref{fig:masslimit}, and in the \FSM case the fraction of disks with radius $> \SI{50}{au}$ increases in the first time steps. This is related to the fact that, while some low mass disks get destroyed, others disks are still expanding due to viscous evolution. Some of these $\mathrm{R}_\mathrm{disk} > \SI{50}{au}$ disks could still have masses or surface densities that are too low to form a planetary system.

\begin{figure}
  \includegraphics[width=\linewidth]{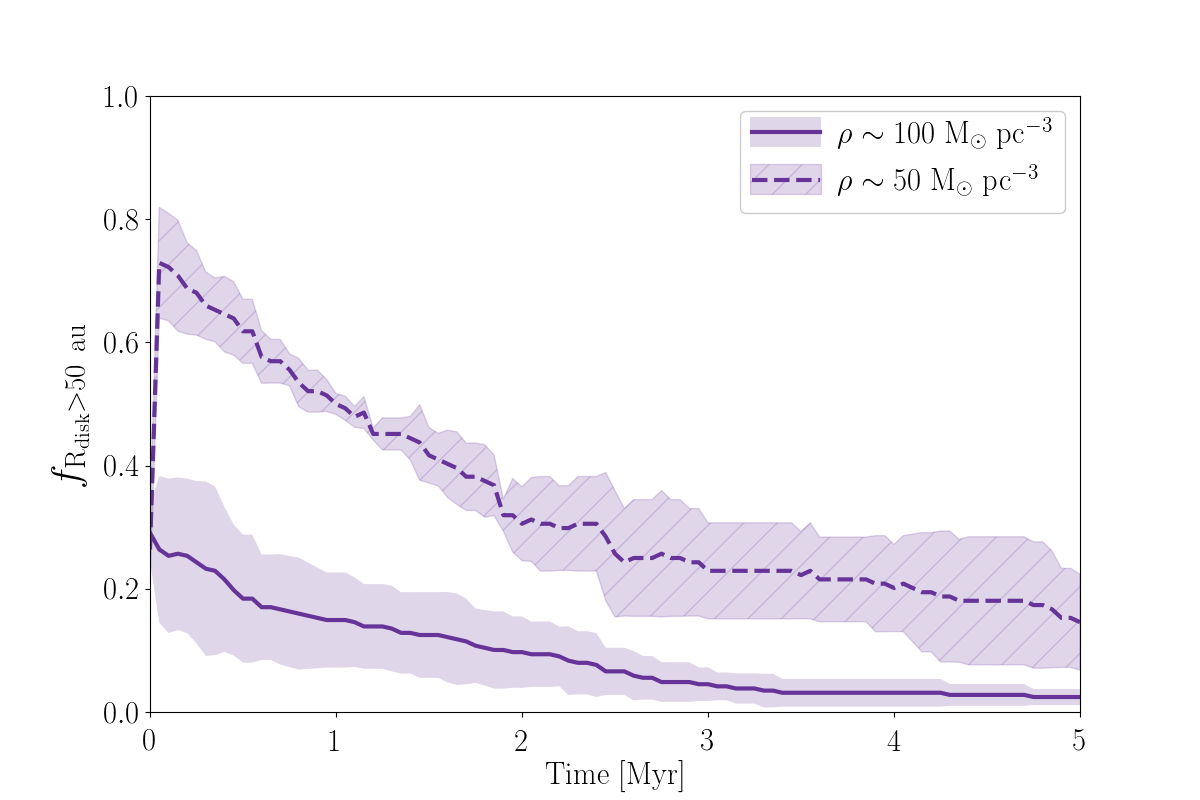}
  \caption{Fraction of disks with radius $\mathrm{R}_{disk} > 50 au$ in time, for regions of different stellar densities. The shaded areas indicate the standard deviation of the simulations.}
  \label{fig:sizelimit}
\end{figure}

\section{Discussion}
\label{sec:discussion}

\subsection{Disk survival and consequences for planet formation}\label{discussion_disksurvival}
The results of the simulations carried out in this work characterize external photoevaporation as an important mechanism for disk dispersion. In comparison, the effect of dynamical truncations is negligible as a means for disk destruction. 

The mean radiation received by the stars in our simulations fluctuates around $\sim 500 \mathrm{\ G}_0$ for the  \HSM region and around $\sim 50 \mathrm{\ G}_0$ for the  \FSM region. The FUV flux in the ONC is estimated to be $\sim 3 \times 10^4 \mathrm{\ G}_0$ \citep{odell1994}. \citet{kim2016} estimate $\sim 3000 \mathrm{\ G}_0$ around a B star in NGC 1977, a region close to the ONC.  According to this and to our results, most of the disks in such a dense region would be destroyed before reaching $\SI{1}{Myr}$ of age. Figure \ref{fig:diskfractions} also agrees with results by \citet{facchini2016} and \citet{haworth2017}, which show that photoevaporation mass loss can be important even for regions with $\sim 30 \mathrm{\ G}_0$ and $\sim 4 \mathrm{\ G}_0$, respectively.  In particular, our results for the  \FSM region show that even very low FUV fields can be effective in dispersing circumstellar disks over time. \citet{winter2019a} find similar dispersion timescales, with a median of $\SI{2.3}{Myr}$ in the solar neighbourhood and $\SI{0.5}{Myr}$ in the central regions of the Milky Way, for stars down to $0.8 \mathrm{M}_\odot$. Comparable results are obtained by \citet{nicholson2019}, who find the half life of protoplanetary disks to be around $2 - \SI{3}{Myr}$ in clusters of various initial conditions.

Protoplanetary disks need to have a reservoir of dust mass $\mathrm{M}_{dust} \gtrsim 10 \mathrm{M}_{\oplus}$ to be able to form the rocky cores of giant gas planets \citep{ansdell2018}. \citet{manara2018} show that such cores need to be already in place at ages $\sim 1 - 3 $ Myr for this type of planets to form. Figure \ref{fig:masslimit} is in agreement with these conclusions. In our simulations, by $\SI{1}{Myr}$ around $20\%$ of the disks have masses $\gtrsim 10 \mathrm{M}_{\oplus}$. This number drops to $\sim 10\%$ by $\SI{3}{Myr}$. According to our results rocky planets and gas giant cores must form very early on, otherwise the protoplanetary disks are not massive enough to provide the necessary amount of solids. This is in agreement with observational time constrains for planet formation and with the so-called ``missing-mass`` problem: solids mass measurements in protoplanetary disks are lower than the observed amount of heavy elements in extrasolar planetary systems around the same type of stars \citep[see e.g.][for discussions on this topic]{manara2018,najita2014,williams2012,greaves2010}. Two scenarios have been proposed to explain this discrepancy in disk and exoplanet masses. The first one suggests that planet cores emerge within the first Myr of disk evolution, or even during the embedded phase while the disk is still being formed \citep[e.g.][]{williams2012,greaves2010}. The second scenario proposes that disks can work as conveyor belts, transporting material from the surrounding interstellar medium towards the central star \citep[e.g.][]{throop2008,kuffmeier2017}.

Disk dispersal is not homogeneous across stellar types. There are observational indications that disk dispersion timescales depend on the mass of the host star, and that less massive ($\sim 0.1 - 1.2 \mathrm{\ M}_\odot$) stars keep their disks for longer than massive stars \citep{carpenter2006b, carpenter2009, luhman2012}. We do not see this effect in our simulations, where the most massive stars keep their disks for longer simply because they initially have the most massive disks. The same effect is observed in \citet{winter2019a}, who used an analytic approach to estimate protoplanetary disk dispersal time scales by external photoevaporation. The discrepancy between observations and theoretical results suggests that internal processes not considered in this work can also play an important role in disk dispersal. Radial drift of dust, fragmentation of large grains, and planetesimal formation are observed mechanisms that can affect both disk lifetimes and observed disk sizes. Viscous evolution alone is another internal process that can contribute to disk dispersal. A more complete model of disk evolution is needed to include the interplay between internal and external dispersion processes.

Initial disk masses are currently highly uncertain. Our chosen value of $\mathrm{M}_d(t=0) = 0.1 \mathrm{M}_* $ is arbitrary, but disks of higher masses could still be stably supported \citep{haworth2018a, nixon2018}.

Once a planetary system has formed, its survival inside a star cluster is not guaranteed. Of the 4071 exoplanets confirmed to date, only 30 have been found inside star clusters. \citet{cai2019} performed simulations of planetary systems in dense, young massive star clusters. They found that the survival rate is $< 50\%$ for planetary systems with eccentricities $e \sim 0.05$ and semi-major axes $<\SI{20}{au}$ over $\SI{100}{Myr}$ of evolution. \citet{vanelteren2019} find that, in regions such as the Trapezium cluster, $\sim 30\%$ of planetary systems are affected by the influence of other stars. Their fractal initial conditions provide local regions of higher densities, which are more favorable for dynamical encounters than our initial conditions. When making parallels with currently observed exoplanet systems, it is important not only to consider the environment effects on the early protoplanetary disks, but also on the planets themselves once they are already formed.

Observations suggest that planets are able to circumvent all of these adversary processes and still form in highly unlikely regions. Evidence of star formation and even proplyd-like objects have been observed around Sgr A* \citep{yusef-zadeh2015, yusef-zadeh2017}. Free-floating planets have been observed in the galactic center, and efficiency analyses of these detections suggest that there are many more yet to be observed \citep[e.g.][]{ryu2019,navarro2019}.

\subsection{Influence of initial conditions}\label{disc:ic}
The effect of switching on photoevaporation when our simulations start have dramatic consequences for the initial circumstellar disks. Mass loss due to photoevaporation occurs very quickly once the stars are immersed in the FUV field. Around $60\%$ and $20\%$ of the disks are dispersed within the initial $\SI{50000}{yr}$ in the \HSM and \FSM regions, respectively. The mean mass of the host stars whose disks dispersed in the initial $\SI{50000}{yr}$ is $0.17 \pm 0.03 \mathrm{M}_\odot$ for the \HSM region and $0.14 \pm 0.04 \mathrm{M}_\odot$ for the \FSM region. In Figure \ref{fig:mass_difference} we show the disk fractions in time, separately for stars with masses $\mathrm{M}_* < 0.5 \mathrm{M}_\odot$ and $\mathrm{M}_* \geq 0.5 \mathrm{M}_\odot$. It can be seen that, for the stars of masses $\mathrm{M}_* < 0.5 \mathrm{M}_\odot$, the disk fractions drop much more dramatically during the first thousand years of cluster evolution.

\begin{figure}
  \includegraphics[width=\linewidth]{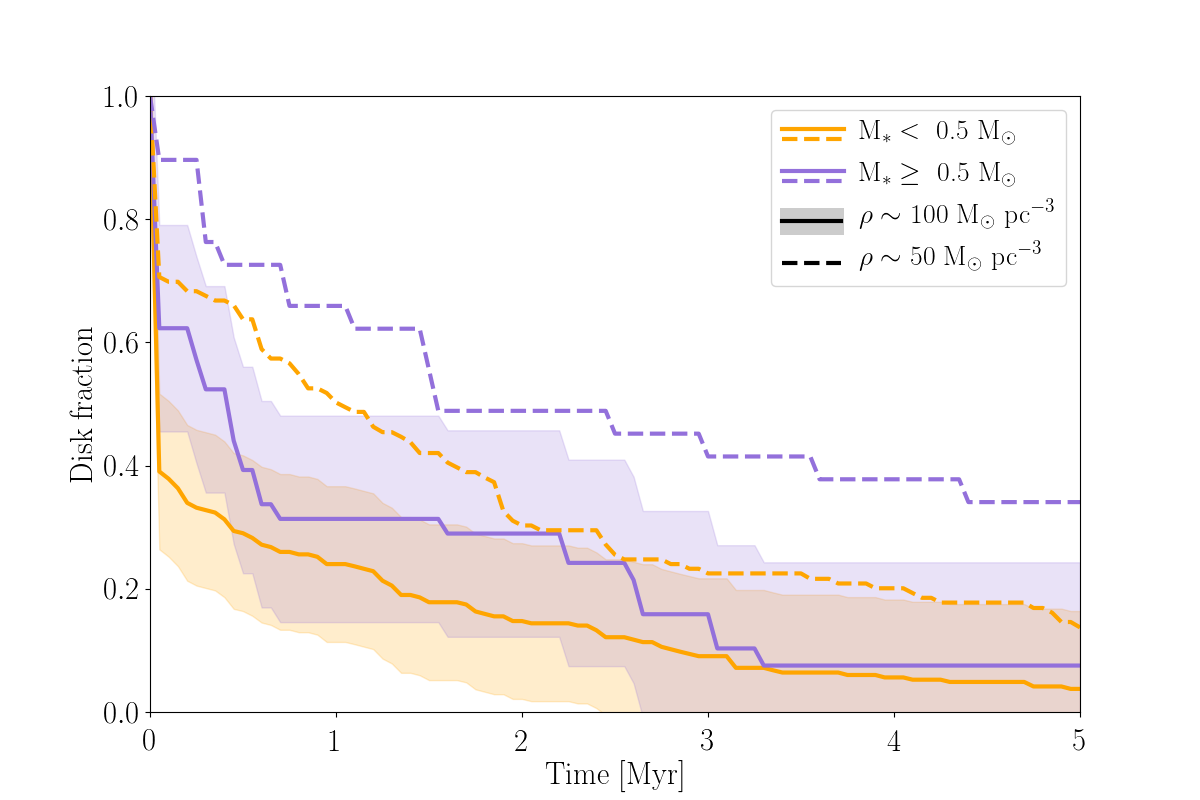}
  \caption{Disk lifetimes for stars $\mathrm{M}_* < 0.5 \mathrm{M}_\odot$ (orange) and $\mathrm{M}_* \geq 0.5 \mathrm{M}_\odot$ (purple). The shaded areas indicate the standard deviation of the simulations. For clarity, only the standard deviation for \HSM is shown, but the one for \FSM is of similar magnitude.}
  \label{fig:mass_difference}
\end{figure}

In reality, if circumstellar disks do form around such low mass stars, they could be sheltered from photoevaporation by interstellar gas and dust, which can linger for several million years after star formation \citep{portegieszwart2010}. Models of the Cygnus OB2 region by \citet{winter2019} demonstrate that the extinction of FUV photons through the gas dampens the mass loss of the disks, increasing their lifetimes. They show that Cygnus OB2 probably underwent a primordial gas expulsion process that ended $\sim \SI{0.5}{Myr}$ ago. This is based on the fact that $\SI{0.5}{Myr}$ of exposure to FUV fields reproduces the observed disk fractions in the region. Given that the estimated age of Cygnus OB2 is $3 - \SI{5}{Myr}$ \citep{wright2010} the primordial gas in the region insulated the disks from external photoevaporation for several Myr. A similar point is made by \citet{clarke2007}, who propose that the FUV field of the star $\theta_1 \mathrm{\ Orionis\ C}$ must have been ``switched on`` no more than $1 - \SI{2}{Myr}$ ago to explain the disk size distribution observed around it. This switching on could have been caused by the star clearing out the primordial gas it was embedded in, thus reducing extinction around it and making its effective radiation field stronger \citep{ali2019}. The presence of gas in young star clusters could then protect the protoplanetary disks and make the disk fraction drop more smoothly than what is shown in Figure \ref{fig:diskfractions}.

The observations shown in Figure \ref{fig:diskfractions} span clusters of many different ages and densities. Our simulation results show that one order of magnitude difference in initial cluster density can yield an important difference in the number of surviving disks. A one order of magnitude spread in cluster density translates to a three order of magnitude difference in cluster radius. The extent of stellar densities in regions where circumstellar disks have been detected does not only reflect the environment of these regions, but also the variety in initial cluster densities.

The initial spatial distribution of the stars in the simulations also plays an important role during the early stages of disk evolution. The stars in our simulations were initially distributed in a Plummer sphere with a specified radius and in virial equilibrium. An approach with fractal or filamentary \citep[e.g.][]{winter2019} initial conditions could change the overall disk survival rates. If a massive star is born in a clump of a fractal distribution, for example, stars in other clumps without massive stars could be minorly affected by radiation and have higher chances of surviving and, eventually, making planets. Higher density regions also increase the relevance of dynamical truncations. This effect of initial conditions could also be counteracted by dynamical mass segregation, in which massive stars move towards the center of the cluster. This would increase the effect of photevaporation in the central regions of the cluster.

\subsection{Model caveats}\label{caveats}
There are several physical processes not considered in this work which could affect the results presented here. One big caveat of our model is the lack of separation between dust and gas components in circumstellar disks. These separate disk components evolve differently and are affected in distinctive ways by outside mechanisms such as the ones implemented in this work. Gas disks has been observed to be larger than dust disks by a factor of $\sim2$ \citep{ansdell2018}. Whether this is caused by different evolution for gas and dust or observational optical depth effects is still up for debate \citep[see e.g.][for discussions on the topic]{birnstiel2013, facchini2017, trapman2019}. The dust in protoplanetary disks is subject to radial drifting and radially dependent grain growth, which can make it resilient to photoevaporation. This can have direct implications on the photoevaporation mass loss rates \citep{facchini2016, haworth2018a} and consequences on planet formation. The conclusions regarding planet formation timescales derived in this work only consider the life expectancy of the disks, but considering different dust and gas disk components will likely affect these results.

While photoevaporation is considered to be primarily damaging for disks when coming from external sources, under certain regimes the photons coming from the host star can also contribute to disk dispersal. \citet{gorti2009} and \citet{gorti2009a} show that FUV photons from the host star can drive photoevaporation mass loss at disk radius $\sim 1 - 10 \mathrm{\ au}$ and $\gtrsim 30 \mathrm{\ au}$. \citet{owen2010} and \citet{font2004} show that internal photoevaporation can also remove loosely bound material from the outer regions, however the largest mass loss was from the inner $\sim\SI{20}{au}$ region. \citet{fatuzzo2008} and \citet{hollenbach2000} find that external photoevaporation dominates for disk regions $> \SI{10}{au}$. The approach used in this work is valid for the disk truncation approximation, however, a more complete analysis would have to consider the combined action and interplay of external and internal photoevaporation.

Mass loss due to photoevaporation was modelled by calculating a truncation radius and removing all the mass outside it, while the inner region of the disk remained unperturbed. In reality, external FUV radiation can heat the whole surface of the disk, and mass loss can occur not just as a radial flow but also as a vertical flow from all over the disk \citep{adams2004}. Given that the mass in the outer regions of a disk is more loosely bound to its host star, the truncation approach is a good first order approximation for mass loss. Furthermore, the FRIED grid used to estimate the photoevaporation mass loss was built using a 1-dimensional disk model. New simulations by \citet{haworth2019} show that, when considering a 2-dimensional disk model, mass loss rates can increase up to a factor 3.7 for a solar mass star. The photoevaporation mass losses obtained in this work should then be considered as lower limits, but are still a good estimate of the effects of bright stars in the vicinity of circumstellar disks.

In the present work we did not include binary stars or any multiples. The presence of multiple stellar systems can have direct consequences on the dynamical evolution of the cluster and on the effects of photoevaporation over the disks. Disks around binary stars have been observed in the star forming regions $\rho$ Ophiuchus \citep{cox2017} and Taurus-Auriga \citep{harris2012,akeson2014,akeson2019}. Observations suggest that these disks are more compact and less bright than the ones around isolated stars. Disks around binary stars might also have shorter lifetimes, due to effects of the companion on the viscous timescale of the disk and also because of photoevaporation inside the system \citep{shadmehri2018,rosotti2018a}.

Another process that can have important consequences in the evolution of circumstellar disks are supernovae explosions. \citet{close2017} showed that nearby ($\SI{0.3}{pc}$) supernova explosions can cause mass loss rates of up to $\SI{1E-5}{M_\odot yr^{-1}}$ which can be sustained for about $\SI{200}{yr}$. Only disks that are faced with the flow face-on manage to survive, but still lose $50\%$ of their mass in the process. \citet{portegieszwart2018} show that a supernova explosion at a distance between $0.15$ and $\SI{0.4}{pc}$ could create a misalignment of $\sim 5\degree.6$ between the star and its disk, which is consistent with the inclination of the plane of the Solar System. Such an event would also truncate a disk at around the edge of the Kuiper belt ($42 - \SI{55}{au}$). Similar effects can be caused by other outcomes of stellar evolution, such as winds \citep{pelupessy2012}.

\section{Conclusions}\label{sec:summary}

We perform simulations of star clusters with stellar densities \rHSM and \rFSM. The stars with masses $\mathrm{M}_* \leq 1.9 \mathrm{\ M}_\odot$ are surrounded by circumstellar disks. Stars with masses $\mathrm{M}_* > 1.9 \mathrm{\ M}_\odot$ are considered sufficiently massive stars to emit UV radiation, causing the disks around nearby stars to evaporate. The disks are subject to viscous growth, dynamical encounters, and external photoevaporation. The simulations span $\SI{5}{Myr}$ of cluster evolution. The main results of this work are:

\begin{enumerate}[leftmargin=*]
    \item[1.] In clusters with density \rHSM around $80\%$ of disks are destroyed by external photoevaporation within $\SI{2}{Myr}$. The mean background FUV field is $\sim 500 \mathrm{\ G}_0$.
    \item[2.] In clusters with density \rFSM around $50\%$ of disks are destroyed by external photoevaporation within $\SI{2}{Myr}$. The mean background FUV field is $\sim 50 \mathrm{\ G}_0$. This shows that even very low FUV fields can be effective at destroying disks over long periods of time.
    \item[3.] Mass loss caused by dynamical encounters is negligible compared to mass loss caused by external photoevaporation. Disk truncations that result from dynamical encounters are not an important process in setting observed disk size and mass distributions.
    \item[4.] At $\SI{1}{Myr}$, $\sim 20\%$ of disks in the \rHSM region and $\sim 50\%$ of disks \rFSM region have masses $\mathrm{M}_{disk} \geq 10 \mathrm{\ M}_\oplus$, the theoretical limit for gas giant core formation.
    \item[5.] Our results support previous estimations that planet formation must start in timescales $< 0.1 - 1 \mathrm{\ Myr}$ \citep[e.g.][]{najita2014, manara2018}.
    \item[6.] The obtained disk fractions in the different density regions, together with the quick dispersion of the disks in all the simulations, suggest that initial conditions are very important in the development of models of early protoplanetary disk evolution.
    \end{enumerate}
    
\section*{Acknowledgements}
We would like to thank the anonymous referee for their thoughtful comments that helped improve this paper. We would also like to thank Andrew Winter, Sierk van Terwisga, and the protoplanetary disk group at Leiden Observatory for helpful discussions and comments. F.C.-R. would like to thank Valeriu Codreanu from SURFsara for his invaluable technical assistance. The simulations performed in this work were carried out in the Cartesius supercomputer, part of the Dutch national supercomputing facility SURFsara. 

\appendix
\section{Resolution of the disks}\label{appendix:resolution}
We use a resolution of 50 cells for the disks which gives us a good trade-off between computing time and acceptable results. In isolated disk evolution this causes an overestimation of disk radius by $\sim 10\%$ on average over 1 Myr of disk evolution, compared with higher resolutions (Figure \ref{fig:radii}). Since all the disks in the simulation are affected by photoevaporation from the start, no disks evolve as in the isolated case. Disk masses are overestimated by less than $\sim 5\%$ compared to higher resolution runs (Figure \ref{fig:cumulative_mass}). This results, in turn, in a slight underestimation of the effects that mass removal, whether through photoevaporation or dynamical encounters, has on the survival times of the disks. Given that we define a disk as dispersed when it has lost $\sim 90\%$ of its initial mass, the slight mass overestimate obtained with the 50 cells resolution does not reflect in a quicker evaporation of the disks.

\begin{figure}
  \includegraphics[width=\linewidth]{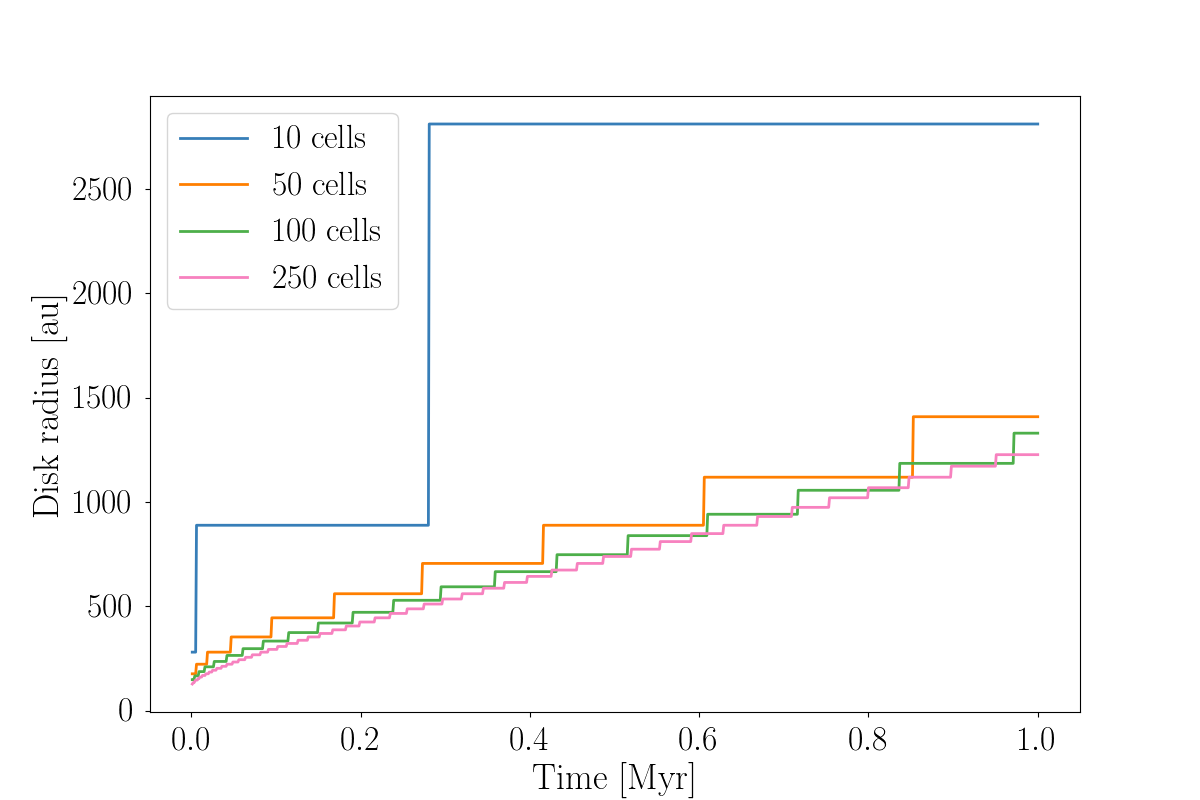}
  \caption{Disk radius in time for different resolutions, for a disk evolving in isolation for $\SI{1}{Myr}$.}
  \label{fig:radii}
\end{figure}

\begin{figure}
  \includegraphics[width=\linewidth]{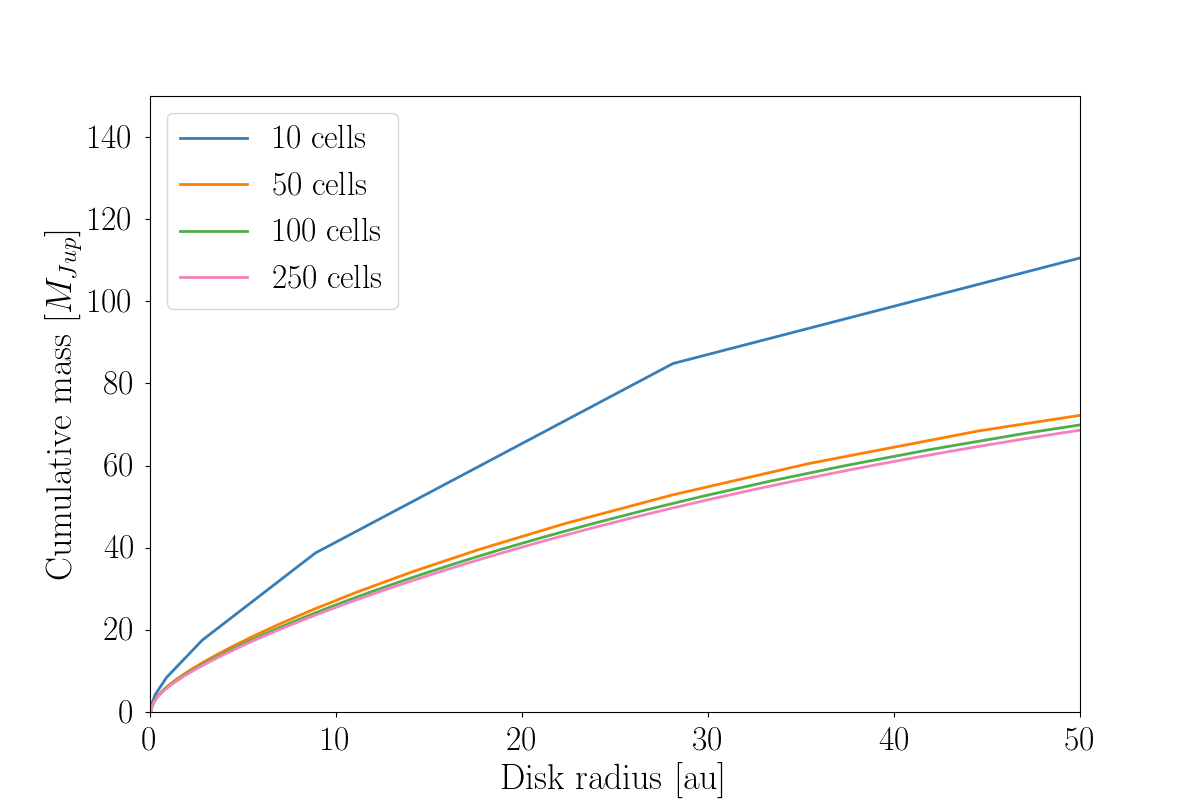}
  \caption{Cumulative disk mass for different resolutions, for a disk evolving in isolation for $\SI{1}{Myr}$.}
  \label{fig:cumulative_mass}
\end{figure}


\bsp	
\bibliographystyle{mnras}
\bibliography{references}
\label{lastpage}
\end{document}